\documentstyle[epsfig,psfig]{texas}

\begin{document}

\title{PREHEATING, PARAMETRIC RESONANCE AND THE EINSTEIN FIELD EQUATIONS}
\author{Matthew PARRY and Richard EASTHER}
\address{Department of Physics, Brown University\\
Box 1843, Providence RI 02912, USA\\
{\rm Email: parry@het.brown.edu, easther@het.brown.edu}}

\begin{abstract}
We consider the interaction between perturbations in the inflaton and in
the metric during the preheating phase in simple inflationary models. By
numerically integrating the Einstein field equations we are able to gauge
the impact of non-linear gravitational effects on preheating for the first
time. In the $\lambda \phi^4$ model we find a large increase in the
amplitude of sub-Hubble metric modes, beyond that due to gravitational
collapse alone. There is significant mode-mode coupling and the
amplification is eventually terminated by back reaction effects. We
suggest that such enhancement of inhomogeneity will change the behaviour
of the post-inflationary universe.

\end{abstract}

\section{Introduction}

Inflationary cosmology proposes a period of nearly exponential growth in
the primordial universe and in so doing solves a number of the problems
that face big bang models. In most implementations of inflation, the
post-inflationary universe is a very cold and very empty place, and the
first question one has is, where did the all matter seen today came from?
The short answer is that it was due to a process called {\it reheating} in
which the inflaton decayed and released its energy into ``normal'' matter
and radiation. It was realised in 1990, however, that the long answer, the
details of reheating, might turn out to be very interesting indeed. 

Traschen and Brandenberger\cite{tb90} showed that if the inflaton field
oscillates coherently at the end of inflation, then it may drive a period
of explosive particle production through the mechanism of parametric
resonance. This process is often termed {\it preheating} and its
ramifications for cosmology include the areas of non-thermal phase
transitions and defect production, baryogenesis, dark matter,
gravitational waves and primordial black holes (see \cite{kls97,pe99,sp98}
and refs. within). In models which allow preheating, {\it re}heating is
now understood as preheating followed by thermalisation.

In the standard analysis of preheating one assumes a
Friedmann-Lema\^{i}tre-Robertson-Walker (FLRW) universe and then
considers, both analytically\cite{kls97,gkls97} and
numerically\cite{kt96,pr97}, the field dynamics in this background. In
general one finds that certain modes in the momentum space of the fields
are exponentially populated, with the precise structure of these {\it
resonance bands} being highly model-dependent. The result is that the
fields soon deviate from homogeneity and there are significant non-linear
effects such as rescattering of the particles, possible symmetry
restoration and back reaction on the inflaton. Although the FLRW scale
factor can be calculated self-consistently\cite{rh97,bcvhss97} in these
scenarios, it was realised recently that the impact of
inhomogeneity\cite{kh96,hk96,nt96,bkm98,fb99,pe99,btkm99,ep99} and
non-linearity\cite{pe99,btkm99,ep99} on the metric must also be
considered. 

Our analysis of preheating is new in that we use the full Einstein field
equations. This enables us to study non-linear gravitational effects that
are not accessible in the perturbative approximation. We find that
parametric resonance can induce large metric inhomogeneities and that
mode-mode coupling and back reaction is significant. At present, our
principal simplifying assumption is that the inhomogeneity is in one
spatial direction only. This means we have a $1+1$-dimensional system of
partial differential equations for two metric functions. 

The larger goal we are pursuing is to determine the spectrum of amplified
fluctuations, in both the fields and the metric, that arise from
preheating. At issue is whether predictions of preheating are in accord
with cosmological observations. Recently it has been
argued\cite{bkm98,btkm99} that super-Hubble modes can be amplified during
preheating. The rationale is that inflation leaves the oscillating
inflaton coherent on scales much larger than Hubble radius. If this is the
case, one might have to reconsider the standard perturbative account of
the fluctuations that lead to large-scale structure formation, especially
if mode-mode coupling occurs. On the other hand, it is certainly the case
that amplified fluctuations on Hubble and sub-Hubble scales will give rise
to gravitational waves\cite{kt97,bkm98}, and one would also like to know
if primordial black hole formation after inflation is
affected\cite{bkm98,ep99}.

Here we briefly survey our progress to date. We have considered preheating
in two models of chaotic inflation, supposing there are no fields other
than the inflaton. We have confirmed\cite{pe99} an analytical
result\cite{fb99} that the $m^2\phi^2$ model does not undergo parametric
resonance. In the $\lambda\phi^4$ model where the standard analysis
predicts parametric resonance in the field we have found strong
amplification of the corresponding metric modes, and coupling of these
modes to modes outside the resonance band. The modes which are amplified
are sub-Hubble modes. We will report more fully on these results in a
forthcoming paper\cite{ep99}. 

\section{Metric and initial conditions}

For simplicity we have assumed a universe with planar symmetry
\footnote{One may think of this as a universe with a flat, 2-dimensional,
maximally symmetric subspace; in contrast the FLRW metric has a
3-dimensional, maximally symmetric subspace.}, so the metric functions
depend only on time and one spatial co-ordinate. We studied preheating
after $m^2\phi^2$ inflation\cite{pe99} using the metric
\begin{equation}\label{met1}
ds^2 = dt^2 - A^2(t,z)\,dz^2 - B^2(t,z) \left ( dx^2 + dy^2 \right ).
\end{equation}
In the limit of small spatial inhomogeneity this resembles a flat FLRW
metric written in terms of physical time. We could also use this metric
for the $\lambda\phi^4$ case, but it turns out to be more convenient to
introduce a ``conformal-like'' metric. This is obtained via a co-ordinate
transformation: $\eta = \eta(t,z)$ and $\zeta = \zeta(t,z)$, which allows
(\ref{met1}) to be written as
\begin{equation}\label{met2}
ds^2 = \alpha^2(\eta,\zeta) \left ( d\eta^2 - d\zeta^2 \right ) -
\beta^2(\eta,\zeta) \left ( dx^2 + dy^2 \right ).
\end{equation}
Here, we will concentrate on this second metric as we have discussed
(\ref{met1}) elsewhere\cite{pe99} and the way we choose the initial
conditions is the same in both cases.

The Einstein field equations, $G^{\mu\nu} = -\kappa T^{\mu\nu}$, give
equations of motion, i.e. equations involving second derivatives in time,
for $\alpha$ and $\beta$. In addition one obtains two constraint equations
which contain only first derivatives in time. The equation of motion for
$\phi$ follows from $T^{\mu\nu}{}_{;\nu}=0$. (We refer the reader to
\cite{ep99} for the explicit form of these rather lengthy equations.) We
numerically integrate the equations of motion using the same techniques of
\cite{pe99} and, in addition, simultaneously compute the co-ordinate
transformation which takes us back to metric (\ref{met1}). This allows us
to determine $A$ and $B$ as well. 

\begin{figure}[tbp]
\centering
\begin{tabular}{cc}
\epsfig{figure=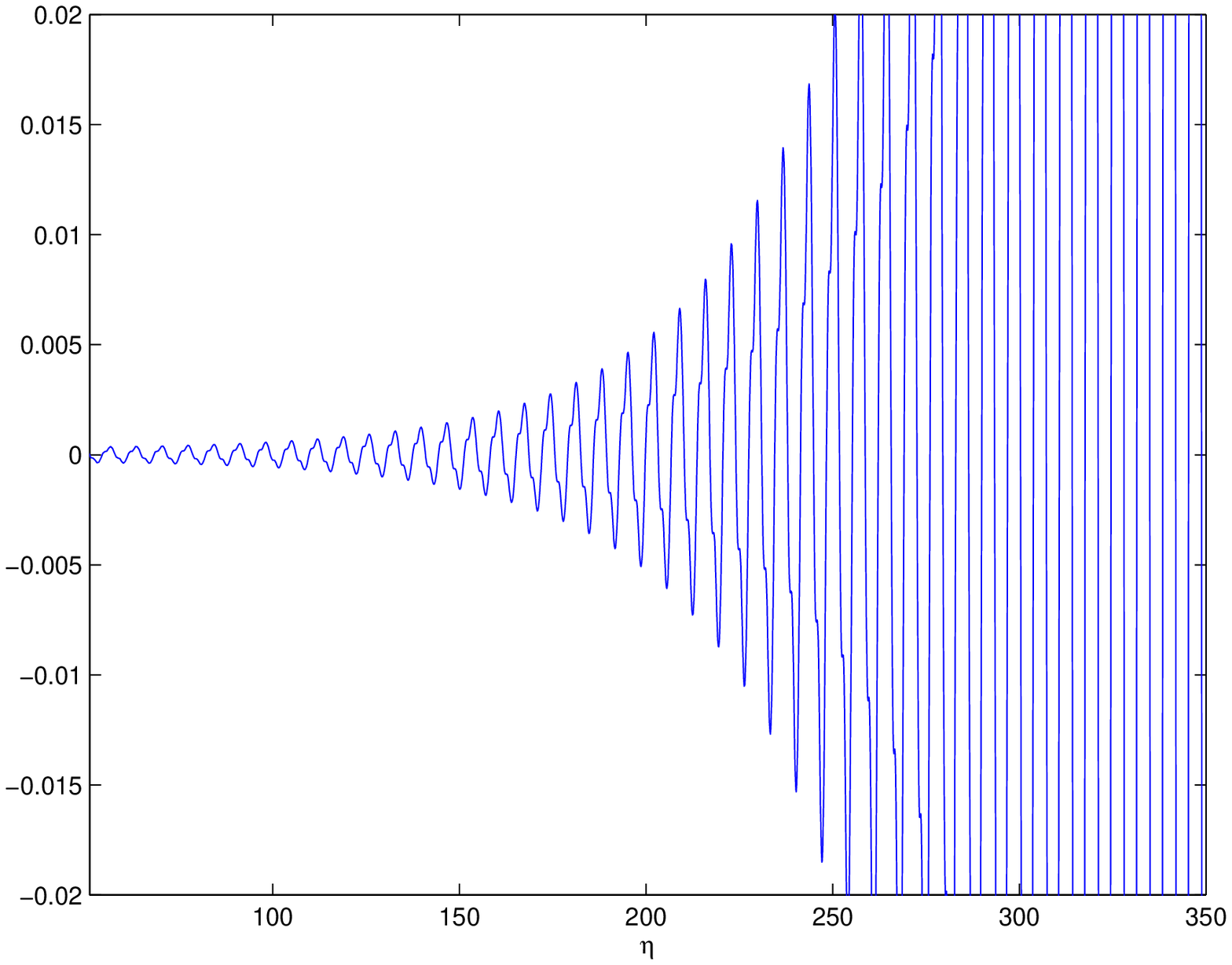,width=7cm} &
\epsfig{figure=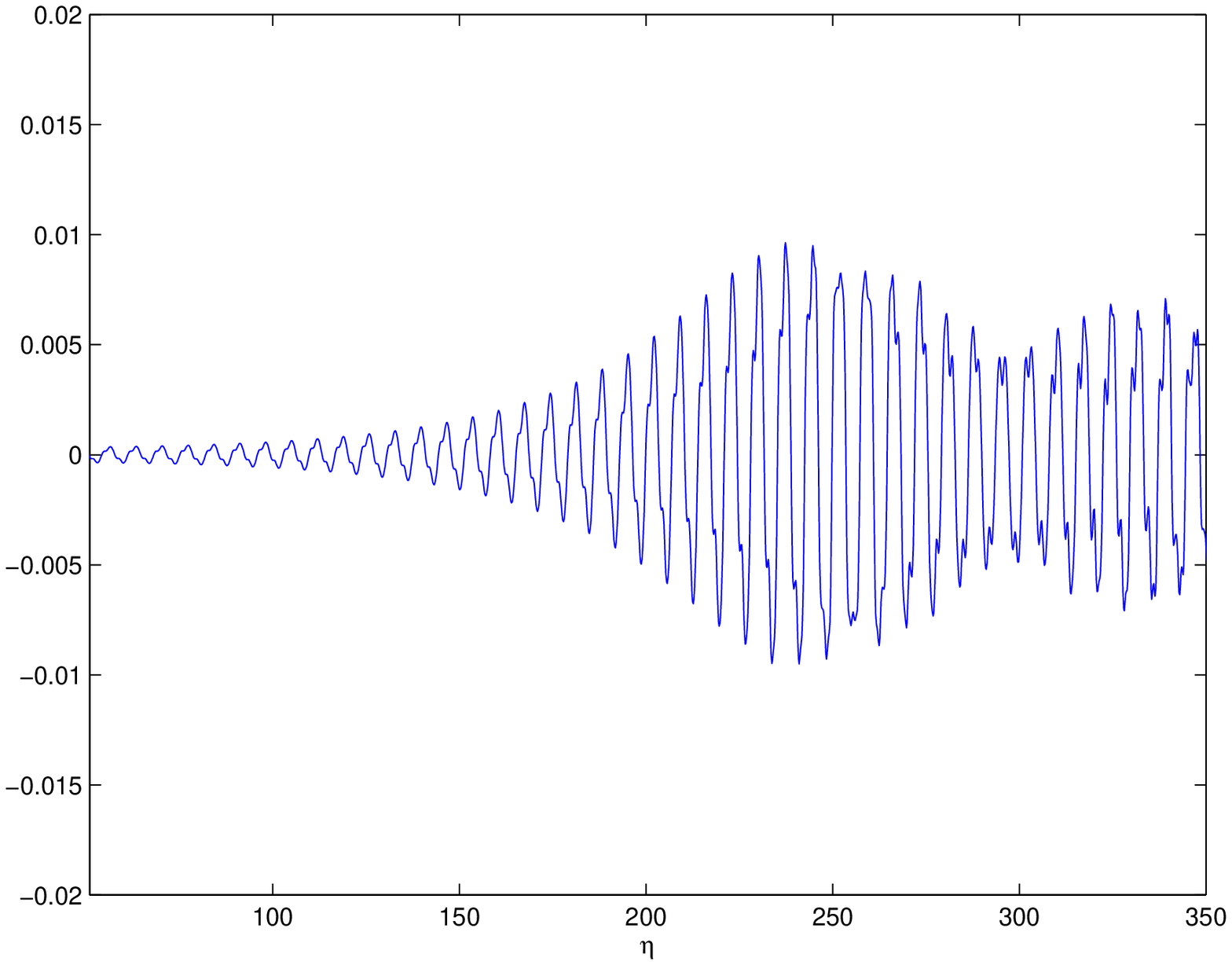,width=7cm}
\end{tabular}
\caption[]{The Fourier transform of $\Phi$, the metric perturbation, for a
resonant mode is shown: the left panel gives the perturbative result,
while the right panel shows the evolution of the mode derived from the
full non-linear analysis.}
\end{figure}

Initial conditions must be chosen to satisfy the constraint equations.
This is, in general, quite difficult to do, but we focus on a relatively
simple scenario where a single mode only is initially excited. Then it is
possible to let $\phi(0,\zeta) = \phi_0$ and $\alpha(0,\zeta) =
\beta(0,\zeta) = 1$. Inhomogeneity is put into the system via the
inflaton's momentum
\begin{equation}
\phi_{,\eta}(0,\zeta) = \dot{\phi}_0 + \epsilon \sin{\left ( {2\pi k \zeta
\over Z} \right ) },
\end{equation}
where $k$ is the number of times the fluctuation will fit into our
simulation ``box'' of length $Z$. The size of the perturbation is governed
by $\epsilon$ and is {\it not} required to be small. We start our
simulations at the end of inflation and this dictates the values of
$\phi_0$ and $\dot{\phi}_0$. Our choice of rescaling makes the initial
density perturbation of order $\epsilon^2$.

The constraint equations then determine that
\begin{eqnarray}
\alpha_{,\eta}(0,\zeta) = {\kappa \over 2C} \left ( {1 \over
2}\,\phi_{,\eta}^2 + V(\phi_0) \right ) - {C \over 2}, \label{ic1} \\
\beta_{,\eta}(0,\zeta) = \sqrt{{\kappa \over 3}\left (
{1 \over 2}\, \langle \phi_{,\eta}^2 \rangle + V(\phi_0) \right )} \equiv C,
\label{ic2}
\end{eqnarray}
where $\langle \cdots \rangle$ denotes a spatial average. In the limit of
no perturbation these initial conditions ensure the metric reverts
to a FLRW type.

\section{Results and discussion} 

In the standard analysis, where one ignores metric perturbations, there is
no parametric resonance in the $m^2\phi^2$ model. We re-examined this
model using our new approach and were able to show that this null result
continues to hold. This was in agreement with a perturbative
analysis\cite{fb99} which served as a check on our method. In addition, we
began to explore the non-linear regime by making the initial perturbations
artificially large. Now mode-mode coupling becomes noticeable and we
conjectured that the effect would be to {\it broaden} the instability
bands in a resonant system. We have been able to confirm this conjecture
in our latest work\cite{ep99}. 

The $\lambda \phi^4$ model is the simplest model which exhibits parametric
resonance and is therefore an important test case for our approach.
Tantalisingly, the modes in resonance are initially just inside the Hubble
radius. The Hubble radius grows subsequently, so we only observe
sub-Hubble mode amplification in this model.

Fig.~1 compares the amplification of a resonant mode in the perturbative
and in the non-linear analysis. To facilitate the comparison we plot the
gauge invariant metric perturbation $\Phi$\cite{mfb92}, which we extract
from our numerical data. The first thing to notice is that the
perturbative analysis of a resonant system always fails; the resonant
growth takes the system away from the perturbative regime. The full
analysis reveals what really happens: resonance terminates due to back
reaction effects. An example of the coupling between the modes that the
perturbative analysis misses can be seen in Fig.~2. 

\begin{figure}[tbp]
\centering
\epsfig{figure=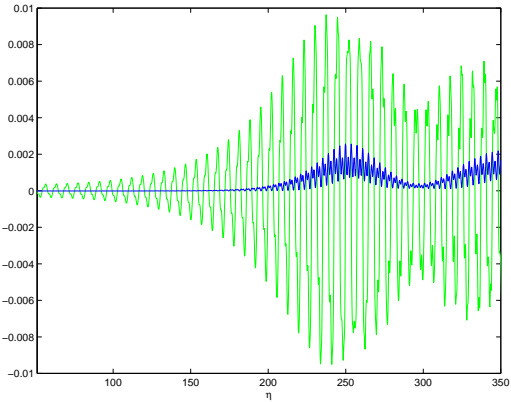,width=7cm} 
\caption[]{The Fourier transform of $\Phi$ for a resonant mode (larger
amplitude) and its second harmonic (smaller amplitude), which is outside
the resonance band. It grows because, in the non-linear analysis, it is
coupled to the resonant mode.}
\end{figure}

Perhaps the most dramatic ramification of resonance can be see in the
evolution of the metric component $A$. When we simulated the effect of an
initial perturbation with wavelength not in the resonance band, we found
only a small increase in the inhomogeneity due to collapse of the
over-dense regions. In the left panel of Fig.~3 one sees the result: $A
\sim \eta$, which is what one expects for a radiation
dominated\footnote{The universe is initially radiation dominated in the
$\lambda\phi^4$ model.}, FLRW universe. The situation was quite different,
however, when we considered a mode in the resonance band. Here the growth
in inhomogeneity is initially {\it driven} by parametric resonance and
leads to a large density contrast, $\delta\rho/\rho \sim 1$. After
parametric resonance terminates, gravitational collapse continues to
enhance the inhomogeneity.

It is because the system is initially driven that we think the formation
of primordial black holes might need to be reconsidered in preheating
scenarios. In the usual understanding, one requires $\delta\rho/\rho \geq
1$ as the mode comes inside the Hubble volume, and black hole formation
proceeds by gravitational collapse. In the $\lambda\phi^4$ model, even
though the relevant metric modes are always sub-Hubble, they are being
amplified resonantly, not gravitationally. 

\begin{figure}[tbp]
\centering
\begin{tabular}{cc}
\epsfig{figure=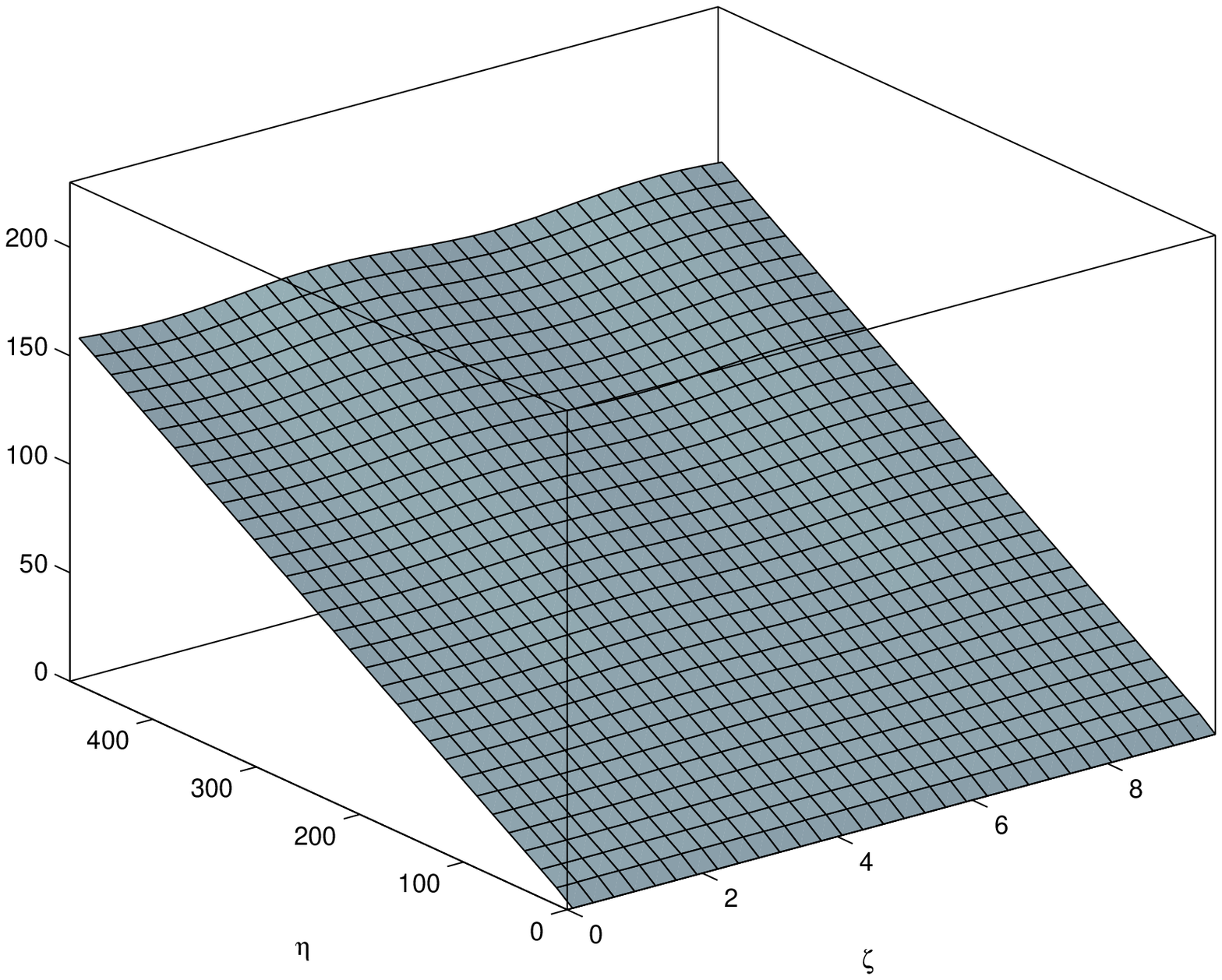,width=7cm} &
\epsfig{figure=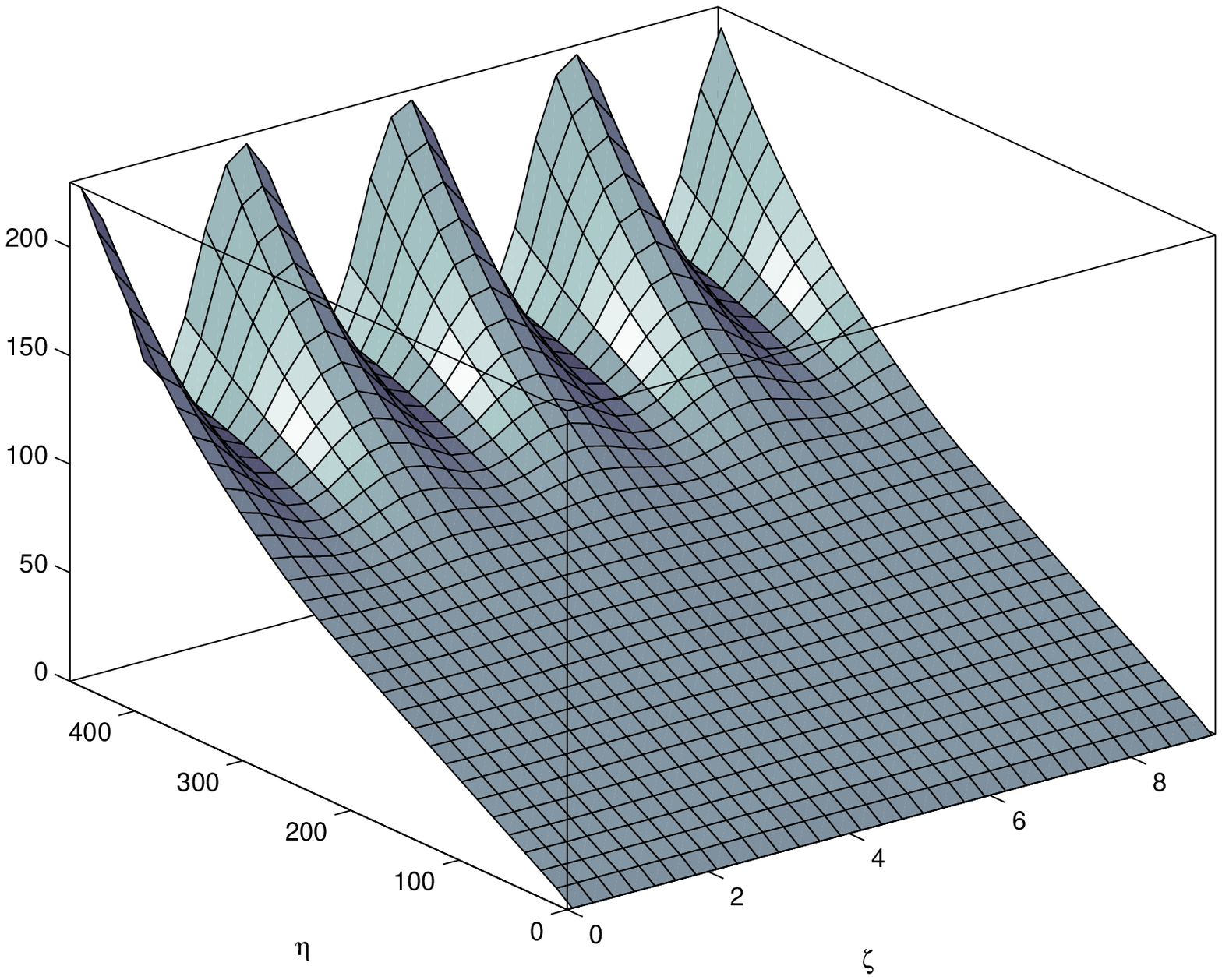,width=7cm}
\end{tabular}
\caption[]{The evolution of $A$ (the $g_{zz}$ component of (\ref{met1})) 
is plotted as a function of $\eta$ and $\zeta$. The left panel shows the
evolution of a mode with $k$ slightly too large to be in resonance, while
in the right panel a mode undergoes resonance, and significant
inhomogeneity is generated.  The units are arbitrary. The initial
perturbation is $\epsilon^2=10^{-6}$, and the simulations begin at the end
of inflation.}
\end{figure}

We are now extending the initial results mentioned above\cite{ep99}. First
we will incorporate general initial conditions and then study more
realistic models which include other fields beside the inflaton. Once we
have determined the spectrum of amplified fluctuations we will be able to
consider whether or not cosmological observations can be used to rule out
certain preheating models.

\section*{Acknowledgments}

We would like to thank Robert Brandenberger, Fabio Finelli, Bruce Bassett
and David Kaiser for useful discussions. Computational work in support of
this research was performed at the Theoretical Physics Computational
Facility at Brown University. RE is supported by DOE contract
DE-FG0291ER40688, Task A.

\end{document}